\documentclass[]{eptcs}
\providecommand{\event}{F-IDE 2021}
\usepackage{breakurl}             
\usepackage{underscore}           
\usepackage{amsmath}
\usepackage{amssymb}
\usepackage{alltt}

\newcommand{\pd}[2]{\dfrac{\partial #1}{\partial #2}}
\newcommand{\tpd}[2]{\frac{\partial #1}{\partial #2}}

\title{A Logic Theory Pattern for Linearized Control Systems%
    \thanks{This research was funded by the Italian Government under the
     CrossLab program, ``Dipartimenti di Eccellenza'' project.}
}
\author{%
Andrea Domenici \qquad\qquad Cinzia Bernardeschi
\institute{Department of Information Engineering\\
University of Pisa\\
Pisa, Italy}
\email{\quad \{andrea.domenici,cinzia.bernardeschi\}@unipi.it}
}

\def\titlerunning{A Logic Theory Pattern}
\def\authorrunning{A. Domenici \& C. Bernardeschi}

\begin{document}
\maketitle

\begin{abstract}
This paper describes a procedure that system developers can follow to
translate typical mathematical representations of linearized control
systems into logic theories.  These theories are then used to verify system
requirements and find constraints on design parameters, with the support
of computer-assisted theorem proving.  This method contributes to the
integration of formal verification methods into the standard model-driven
development processes for control systems.  The theories obtained through
its application comprise a set of assumptions that the system equations must
satisfy, and a translation of the equations into the logic language of the
Prototype Verification System theorem-proving environment.  The method is
illustrated with a standard case study from control theory.
\end{abstract}

\section{Introduction}

The design of modern control systems relies both on theory and on experiment.
Experiment often takes the form of simulation, in a process known as
\emph{model-driven development} (MDD).  Both the theory- and the
experiment-based activities start from a mathematical system model, but in one
case the model is a static subject of formal analysis, and in the other one it
is an active system under observation.  This implies that developers involved
in the two branches of development require different kinds of knowledge and
the model itself may have to be expressed with different notations.

The present paper is an initial contribution to a better integration between
formal analysis and MDD, as it provides a systematic way to translate a
mathematical system description into a logic theory that can be analyzed
with the support of an interactive theorem prover.  The proposed procedure
is tailored to basic linearized control systems and it follows a high-level
pattern, so that its application requires some creativity on the part of
the developers.  The procedure supports developers in providing them with
a framework upon which to build system-specific theories.  Parts of the
procedure can be embodied into code templates for specific classes of systems.
This work also tries to contribute in adding support for control system
development to the large body of theories for the Prototype Verification
System (PVS)~\cite{owre96}.  In particular, a solution for a technical
problem related to the PVS type system is proposed (Sect.~\ref{sm}
and~\ref{mc}).

The PVS environment is introduced in Sect.~\ref{pvs}, Sect.~\ref{pend}
describes the system used as an example, Sect.~\ref{thy} describes the
procedure within the example, and the procedure is exposed in general terms
in Sect.~\ref{gp}.  Conclusions are drawn in Sect.~\ref{lausdeo}.

\section{The PVS theorem-proving environment}
\label{pvs}

The PVS is an interactive theorem prover for theories in higher-order logic.
Theories are written in a rich, strongly-typed language.  In this paper,
only the features needed to understand the examples are introduced.

The type system rests on the fundamental concepts of \emph{reals},
\emph{rationals}, \emph{integers}, etc., whose properties and operations are
formally defined in the built-in theories of the prover.  Subtypes (e.g.,
\texttt{posreal}, positive reals) and  types of arbitrary complexity can be
defined upon this basis.  In particular, \emph{function types} are defined
as in this example:
\begin{small}
\begin{alltt}
phi: VAR [real -> real]
\end{alltt}
\end{small}
where \texttt{phi} is defined as a \emph{function variable} ranging over the
set of functions from reals to reals.  A \emph{function constant} (not to be
confused with a constant function) is defined, e.g., as
\begin{small}
\begin{alltt}
x: VAR real
f(x): real = -(g/R)*sin(x)
\end{alltt}
\end{small}
\emph{Formulae} are made of \emph{predicates} (functions returning a Boolean
value), logical connectives, and quantifiers.  A \texttt{LET ... IN} clause
introduces a definition into the following expression.  A \texttt{COND}
expression selects one value from a set, associated with one of a set of
mutually exclusive Boolean expressions.

Several theories from the NASALIB package~\cite{dutertre96} are
used in this work, in particular, those related to mathematical analysis
and vectors.  Type \texttt{Vect2} represents the set of real 2-vectors.
Function \texttt{vect2} is a constructor, and \texttt{p(0)} (or \texttt{p(1)})
is the first (second) element of a vector \texttt{p}.

Logical statements to be proved are expressed as logic formulae introduced by
a label and a keyword such as \texttt{LEMMA} or \texttt{THEOREM}:
\begin{small}
\begin{alltt}
label: LEMMA f(phi, theta) = g(theta)
\end{alltt}
\end{small}

In the interactive theorem prover, the user selects one lemma as the initial
goal and applies prover commands to transform it.  The commands are based
on the inference rules of sequent calculus~\cite{smullyan68} and perform
transformations of various complexity.  In particular, they may recursively
split a goal into subgoals.  A proof terminates successfully when all subgoals
are satisfied.

\section{Equilibrium of a pendulum}
\label{pend}

A rigid pendulum with a fixed pivot is a simple example of a second-order
dynamical system~\cite{slotine91}.  A pendulum consists in a bob of mass M
and a rod of length R.  The pivot has a coefficient b of viscous friction.
The pendulum is controlled with a driving torque $\tau$ applied at the
pivot.  The dynamic model of the system is then
\[ \mathrm{I}\ddot{\theta}
+ \mathrm{b}\dot{\theta} + \mathrm{M}\mathrm{g}\mathrm{R}\sin(\theta) =
\tau \; , \]
where $\theta$ is the angular displacement from the vertical axis,
$\mathrm{I} = \mathrm{M}{\mathrm{R}}^2$ is the rotational inertia, and
$\mathrm{g}$ is the gravitational acceleration.
Taking $\theta$ and $\dot{\theta}$ as the state variables, the state-space
representation for the free evolution of the system is
\[
\left\{
\begin{aligned}
\dot{x_1} &= \dot{\theta} \\
\dot{x_2} &=  -(\mathrm{g}/\mathrm{R})\sin(\theta)
            - (\mathrm{b}/\mathrm{I})\dot{\theta} \; , \\
\end{aligned}
\right.
\]
where the right-hand members of the equalities are the system's \emph{generating
functions} (GF).  The equilibrium points are $\mathrm{P}_1 = (0, 0)$ and
$\mathrm{P}_2 = (\pi, 0)$.

\section{A logic theory}
\label{thy}

This section shows how a theory can be developed for the formal study of a
dynamical system such as the one introduced above.

In a standard approach, the system is linearized near an equilibrium point,
i.e., its GFs ($f_1$ and $f_2$) are replaced by their differentials.  The
linearized system is expressed as

\[
\dot{\mathbf{x}} = \mathbf{J} \mathbf{x} \: ,
\]
where $\mathbf{x}$ is the state vector and the Jacobian matrix $\mathbf{J}$ is
defined as
\begin{equation*}
\setlength\arraycolsep{5pt}
  \mathbf{J} = \begin{bmatrix}
        \pd{f_1}{\theta} & {} \pd{f_1}{\dot\theta} \\
        \pd{f_2}{\theta} & {} \pd{f_2}{\dot\theta} \\
        \end{bmatrix} \; .
\end{equation*}

\subsection{System model}
\label{sm}

The first step in building the system's theory is then to express the
GFs and their partial derivatives (PD) in the PVS language.
Translating mathematical formulae into PVS is obviously straightforward, but
the theory must provide the means to verify the correctness of the procedure.
In particular, it is necessary to verify that the functions are actually
differentiable, and that the derivatives have been calculated correctly.
Dealing with these issues leads to coping with some limitations in the available
PVS libraries.  In fact, to the best of the authors' knowledge, there are
currently no PVS theories available about partial derivation.  However, a
rich collection of theories about functions of a single variable is available,
which can be used as a basis for a treatment of multi-variable functions.

This work takes a pragmatic approach: Instead of trying to build a general
theory of multi-variable functions, \emph{ad hoc} theories are developed for
specific systems, trying to strike a satisfactory compromise between generality
(and elegance) and synthesis.


From calculus textbooks, the PD of a function $f: \mathbb{R}^n
\rightarrow \mathbb{R}$ wrt a variable $x_i$ in a point $\mathrm{P}_0 = (x_{0
1},\ldots, x_{0 n})$ is the ordinary derivative of the \emph{restriction}
$\phi: \mathbb{R} \rightarrow \mathbb{R}$ of $f$ to the set $\{x_{0 1},\ldots,
x_i,\ldots, x_{0 n}\}$.  This is usually translated as ``\emph{take
the derivative of f wrt $x_i$, keeping the other variables constant}''.
Unfortunately, the PVS type checker ``does not know'' how to keep
variables constant: A variable is supposed to be variable and a constant
is supposed to be constant.  More precisely, it is not possible to treat a
multi-variable function as a single-variable one, take its single-variable
derivative and turn it into a multi-variable function.


In the approach proposed in this work, a developer must write the GFs, the
points where the system is linearized, the single-variable restrictions,
the (ordinary) derivatives of the restrictions, and the PDs of the GFs.
With an explicit definition of the single-variable restrictions and their
derivatives, it is possible to check the correctness of the ``hidden step''
in the common calculation of PDs.  The pendulum system can then be defined as
in the following.  First, parameters, variables, GFs, and equilibrium points:

\begin{small}
\begin{alltt}
M, R, b: posreal        % mass, length, viscous friction
g: posreal              % (standard) gravitational acceleration
I: posreal = M*sq(R)    % rotational inertia
tau: VAR real           % applied torque
theta, dtheta: VAR real % (state variables) angular displmt., time derivative

% generating functions
p: VAR Vect2            % a generic point p = (theta, dtheta)
f1(p): real = p(1)                           % p(1) = dtheta
f2(p): real = -(g/R)*sin(p(0)) - (b/I)*p(1)  % p(0) = theta

% equilibrium points
P1: Vect2 = vect2(0, 0);  X1: real = P1(0); Y1: real = P1(1)
P2: Vect2 = vect2(pi, 0); X2: real = P2(0); Y2: real = P2(1)
\end{alltt}
\end{small}
Then the restrictions are defined by applying the GFs to
points with one variable and one arbitrary constant co-ordinate, and their
derivatives are computed:
\begin{small}
\begin{alltt}
X, Y: real
P: Vect2 = vect2(X, Y)                       % an arbitrary constant point
phi1_x(theta): real = f1(vect2(theta, Y))    % restr. of f1 wrt theta
phi1_y(dtheta): real = f1(vect2(X, dtheta))  %        "         dtheta
phi2_x(theta): real = f2(vect2(theta, Y))    % restr. of f2 wrt theta
phi2_y(dtheta): real = f2(vect2(X, dtheta))  %        "         dtheta

% derivatives of the restrictions
dphi1_x(theta): real = 0;                 dphi1_y(dtheta): real = 1
dphi2_x(theta): real = -(g/R)*cos(theta); dphi2_y(dtheta): real = -(b/I)
\end{alltt}
\end{small}
Finally, the PDs of the GFs are declared:
\begin{small}
\begin{alltt}
df1_dx(p): real = 0                  % partial derivative of f1 wrt theta
df1_dy(p): real = 1                  % partial derivative of f1 wrt dtheta
df2_dx(p): real = -(g/R)*cos(p(0))   % partial derivative of f2 wrt theta
df2_dy(p): real = -(b/I)             % partial derivative of f2 wrt dtheta
\end{alltt}
\end{small}
By convention, letters $x$ and $y$ refer to the
first and second state variable ($\theta$ and $\dot \theta$), respectively.


\subsection{Model consistency}
\label{mc}

In order to verify the correctness of ordinary differentiations, the NASALIB
theories are used.  For example, the following lemma states that the
\texttt{dphi1\_x} is indeed the derivative of \texttt{phi1\_x} wrt
\texttt{theta}, using function \texttt{deriv} defined in the NASALIB
\texttt{derivative} theory:
\begin{small}
\begin{alltt}
der1theta: LEMMA deriv(phi1_x, theta) = dphi1_x(theta)
\end{alltt}
\end{small}

Then the PDs must be proved to be extensions of the derivatives
of the GFs' restrictions.  Predicates \texttt{xrestricts?}
check if a function $\phi: \mathbb{R} \rightarrow \mathbb{R}$ is a restriction
of a function $f:\mathbb{R}^2 \rightarrow \mathbb{R}$ (and conversely).
\begin{small}
\begin{alltt}
x, y: VAR real
f: VAR [Vect2 -> real]        % a generic f: (R x R) -> R
phi: VAR [real -> real]       % its restriction phi: R -> R

xrestricts?(phi, f, p): bool = FORALL (x): phi(x) = f(vect2(x, p(1)))
yrestricts?(phi, f, p): bool = FORALL (y): phi(y) = f(vect2(p(0), y))
\end{alltt}
\end{small}

These predicates are then used first to show that \texttt{phi1_x} etc. are
restrictions of the respective GFs, and then that the
PDs \texttt{df1\_dx} etc. are extensions of the derivatives
of the restrictions.  For example:
\begin{small}
\begin{alltt}
restr1x1: LEMMA xrestricts?(phi1_x, f1, P)
extens1: LEMMA
    xrestricts?(dphi1_x, df1_dx, P) AND yrestricts?(dphi1_y, df1_dy, P)
\end{alltt}
\end{small}

After the consistency of the various functions written by the developer
has been verified, the task remains to check that linearizing the system
at the equilibrium points is mathematically sound, i.e., that the GFs are
differentiable.  Elementary calculus provides a sufficient condition for
differentiability: ``\emph{If f has a partial derivative in a point $P_0$ and
the other partial derivatives exist in a ball around $P_0$ and are continuous
in $P_0$, then $f$ is differentiable in $P_0$}''.  These conditions can be
checked with NASALIB theories on ordinary differentiation and on continuity
in Euclidean spaces. e.g.:
\begin{small}
\begin{alltt}
% does a partial derivative of f1 exist in P1?
exist_pd1P1?: bool = derivable?(phi1_x, X1) OR derivable?(phi1_y, Y1)

% is the partial derivative of f1 continuous at P1?
continuous_pdx1P1?(S: VAR set[Vect2]): bool = continuous_at?(df1_dx, X1, P1)
\end{alltt}
\end{small}

\subsection{Linearization}
\label{lin}

Finally, the Jacobian and its trace and determinant can be defined at the
points of equilibrium.  The \emph{matrices} NASALIB theory was not used as
deemed unnecessarily complex.
\begin{small}
\begin{alltt}
JP1(i, j: below(2)): real =                              % Jacobian at P1
    LET idx = 3*(i - 1) + j IN COND
        idx = 1 -> df1_dx(P1), idx = 2 -> df1_dy(P1),
        idx = 3 -> df2_dx(P1), idx = 4 -> df2_dy(P1)
    ENDCOND

trJP1: real = JP1(1, 1) + JP1(2, 2)                      % trace of JP1

detJP1: real = JP1(1, 1)*JP1(2, 2) - JP1(1, 2)*JP1(2, 1) % determinant of JP1
\end{alltt}
\end{small}
Then, the characteristic polynomial, its discriminant, and the eigenvalues:
\begin{small}
\begin{alltt}
lam: VAR complex
csq(lam): complex = lam*lam    % complex square
charpolJP1(lam): complex = csq(lam) - trJP1*lam + detJP1

% discriminant
discrJP1: real = discr(1, -trJP1, detJP1)

% eigenvalues
lam1: complex =
    IF (discrJP1 >= 0) THEN root(1, -trJP1, detJP1, -1)
    ELSE trJP1/2 - i*sqrt(-discrJP1)/2
    ENDIF
lam2: complex =
    IF (discrJP1 >= 0) THEN root(1, -trJP1, detJP1, 1)
    ELSE trJP1/2 + i*sqrt(-discrJP1)/2
    ENDIF
\end{alltt}
\end{small}
This code has a regular structure and has very few dependencies on the
developer-provided code, so it could be easily turned into a template.

\subsection{Analysis}
\label{an}

Within this theory, it is possible to prove properties of the system, and, in
particular, to find constraints on the physical parameters.  For example, it
has been proved that the system has non-oscillating solutions in $\mathrm{P}_1$
(i.e., the discriminant of the characteristic polynomial is positive) if
and only if the ratio of viscous friction to rotational inertia is greater
than four:
\begin{small}
\begin{alltt}
lem4: LEMMA K = b/I AND K > 4 IFF discrJP1 > 0
\end{alltt}
\end{small}

The proof with the PVS proof assistant is straightforward, involving only
simple manipulations, such as introducing previously proved lemmas and
expanding definitions.  Similarly, it has been proved that $\mathrm{P}_2$ is
unstable.

\section{General procedure}
\label{gp}

The above process can be described in a more general and schematic procedure
as shown below.  It may be noted that, after a developer has written the
system model in PVS, most of the remaining theory, e.g., the lemmas in
Sec.~\ref{mc} and the definitions in Sec.~\ref{lin}, is a set of boilerplate
definitions that could be generated with a template-processing software.

\subsubsection*{System model}

\vspace{-0.5\baselineskip}
The developer defines
(a) the \emph{state variables} $x_1,\ldots, x_n$
(\texttt{theta} and \texttt{dtheta} in Sec.~\ref{sm}),
(b) the \emph{system parameters} $k_1,\ldots, k_l$ (as constants, e.g.,
\texttt{M}, \texttt{R}, \texttt{b}, \texttt{g}),
(c) the system \emph{generating functions} $f_1,\ldots, f_n$
(\texttt{f1}, \texttt{f2}),
(d) the \emph{equilibrium points} (\texttt{P1}, \texttt{P2}),
(e) the \emph{restrictions} $\phi_{1 1},\ldots, \phi_{n n}$
(\texttt{phi1\_x}, \ldots \texttt{phi2\_y}) of the GFs,
(f) the \emph{derivatives} $\phi'_{1 1},\ldots, \phi'_{n n}$
(\texttt{dphi1\_x}, \ldots \texttt{dphi2\_y}) \emph{of the restrictions}, and
(g) the \emph{partial derivatives} $\tpd{f_1}{x_1},\ldots, \tpd{f_n}{x_n}$
(\texttt{df1\_dx}, \ldots \texttt{df2\_dy}) of the GFs.

\subsubsection*{Model consistency}

\vspace{-0.5\baselineskip}
The theory contains predicates and lemmas to check that (a) the $\phi_{i j}$'s
are actually the restrictions of the $f_i$'s, (b) the $\phi'_{i j}$'s are
actually the derivatives of the $\phi_{i j}$'s, (c) the $\tpd{f_i}{x_j}$'s
are extensions of the $\phi_{i j}$'s, and (d) the generating functions are
differentiable, as shown in Sec.~\ref{mc}.  In this phase, lemmas such as
\texttt{restr1x1} and \texttt{extens1} formalize the concept of PD in a way
that is acceptable to a higher-order logic type checker.

\subsubsection*{Linearization}

\vspace{-0.5\baselineskip}
\textbf{For each equilibrium point}, write
(a) the \emph{Jacobian} matrix (\texttt{JP1} in Sec.~\ref{lin}),
(b) the functions of the Jacobian needed to write the characteristic
polynomial, e.g., \emph{trace} and \emph{determinant}
(\texttt{trJP1} and \texttt{detJP1})  for second-order systems),
(c) the characteristic polynomial (\texttt{charpolJP1}),
(d) functions of the polynomial needed to
characterize the set of eigenvalues, e.g., the \emph{discriminant}
(\texttt{discrJP1}) for second-order systems), and
(e) the expressions of the eigenvalues (\texttt{lam1}, \texttt{lam2}).

\subsubsection*{Analysis}

\vspace{-0.5\baselineskip}
Use the functions from the linearization phase to write lemmas about system
properties, e.g., Lemma~\texttt{lem4} in Sec.~\ref{an}, relating stability to
parameter ranges.  In real life applications, this part will require most of
the total effort.  It should be observed that a PVS theory need not be
monolithic, so that the \emph{divide and conquer} principle can and should
applied to cope with problem size and complexity.  Also, formal analysis
requires specific expertise, but a systems engineer can easily learn the
essentials to define the system model and its requirements, leaving theorem
proving to specialized developers (and their software).

\section{Conclusions and further work}
\label{lausdeo}

The procedure proposed in this paper had been used, before being laid down
explicitly, in the analysis of a simple robotic vehicle~\cite{a18} and of a
synchronous motor~\cite{a20a}.  In both cases, formal verification was
complemented by simulation, and in the latter, by design space analysis.  More
precisely, formal verification had been used to find useful ranges of
controller gains and design space analysis, supported by the simulation
environment, was used to find optimal values within those ranges.  The present
work sketches a systematic way to deal with this kind of tasks.  Clearly, the
procedure shown in Sect.~\ref{gp} needs to be defined more in detail and
templates for theory fragments have to be defined.  Also, the analysis of
systems with a large state space will require more advanced strategies to
develop appropriate theories.

\bibliographystyle{eptcs}
\bibliography{f-ide2021}

\end{document}